\documentclass[aps,prb,twocolumn,superscriptaddress]{revtex4}
\usepackage{graphicx}
\usepackage{bm}
\usepackage{amsmath,amssymb,amsfonts,bm}
\usepackage{epsfig}
\usepackage{epstopdf}
\usepackage{dcolumn}
\usepackage{grffile}
\usepackage{verbatim}
\usepackage{mathrsfs}
\usepackage{appendix}
\usepackage[colorlinks=true,linkcolor=blue,citecolor=blue, urlcolor=blue]{hyperref}

\begin{document}


\title{Pristine Mott Insulator from an Exactly Solvable Spin-$1/2$ Kitaev Model}

\author{Jian-Jian Miao}
\thanks{These authors contribute equally.}
\affiliation{Kavli Institute for Theoretical Sciences, University of Chinese Academy of Sciences, Beijing 100190, China}

\author{Hui-Ke Jin}
\thanks{These authors contribute equally.}
\affiliation{Department of Physics, Zhejiang University, Hangzhou 310027, China}

\author{Fa Wang}
\affiliation{International Center for Quantum Materials, School of Physics, Peking University, Beijing 100871, China}
\affiliation{Collaborative Innovation Center of Quantum Matter, Beijing 100871, China}

\author{Fu-Chun Zhang}
\affiliation{Kavli Institute for Theoretical Sciences, University of Chinese Academy of Sciences, Beijing 100190, China}
\affiliation{CAS Center for Excellence in Topological Quantum Computation, University of Chinese Academy of Sciences, Beijing 100190, China}
\affiliation{Collaborative Innovation Center of Advanced Microstructures, Nanjing University, Nanjing 210093, China}

\author{Yi Zhou}
\email{yizhou@zju.edu.cn}
\affiliation{Department of Physics, Zhejiang University, Hangzhou 310027, China}
\affiliation{CAS Center for Excellence in Topological Quantum Computation, University of Chinese Academy of Sciences, Beijing 100190, China}
\affiliation{Collaborative Innovation Center of Advanced Microstructures, Nanjing University, Nanjing 210093, China}

\date{\today}

\begin{abstract}

We propose an exactly solvable quantum spin-$1/2$ model with time reversal invariance on a two dimensional brick-wall lattice, where each unit cell consists of three sites.
We find that the ground states are algebraic quantum spin liquid states. The spinon excitations are gapless
and the energy dispersion is linear around two Dirac points. The ground states are of three-fold topological degeneracy on a torus.
Breaking the time reversal symmetry opens a bulk energy gap and the $Z_2$ vortices obey non-Abelian statistics.

\end{abstract}

\maketitle


\section{Introduction}

Quantum spin liquid (QSL) is defined as a pristine Mott insulator which carries an odd number of electrons per unit cell and hosts paramagnetic ground states\cite{Anderson73,Lee08,Balents10,QSLRMP}.
In such a quantum paramagnet, spontaneous symmetry breaking does not take place because of strong quantum fluctuations, resulting in vanishing entropy density in the thermodynamic limit.
The absence of magnetic ordering and accompanied low-lying magnon excitations in a Mott quantum paramagnet, say, QSL,
offers unique insight into the nature of the intrinsic Mott state.
Thus, searching for QSLs in dimensions greater than one ($D>1$) attracts more and more attentions in both experimental and theoretical sides.
Experimentally, the first QSL was sought in organic salt $\kappa - $(ET)$_2$Cu$_2$(CN)$_3$ on an anisotropic
triangular lattice\cite{Kanoda03},
about 30 years after the proposal of resonating valence bond (RVB) states\cite{Anderson73}.
Since then, several QSL candidates were successively reported in other two quasi-triangular organic compounds,
$[$Pd(dmit)$_2$$]$$_2$(EtMe$_3$Sb)\cite{Itou08} and $\kappa - $(ET)$_2$Ag$_2$(CN)$_3$ \cite{Shimizu16},
the kagome  herbertsmithite
ZnCu$_3$(OH)$_6$Cl$_2$\cite{Kagome07}, and the three-dimensional ($3D$) hyperkagome spinel oxide
Na$_4$Ir$_3$O$_8$\cite{Takagi07} and other emerging materials\cite{QSLRMP}.
Theoretically, a reliable and systematic mathematical framework to understand QSLs is still lacking, although significant progresses toward this aim have been achieved\cite{QSLRMP,Balents17}.
Except for a few rigorous results, we rely heavily on a combination of sophisticated numerical and analytical techniques to ``guess" the ground states and construct the corresponding low-energy excitations.
These rare rigorous results, including exactly solvable models and mathematical theorems, serve as benchmarks in quantum many-body problems. 

The Kitaev quantum paramagnet, represented by an exactly solvable spin-$1/2$ model on honeycomb lattice, was initially invented to illustrate the basic ideas of topological quantum computation\cite{Kitaev},
where non-Abelian anyons are employed to build a fault-tolerant quantum computer\cite{TQCRMP}.
Besides numerous activities motivated by the aspect of fault tolerance, this model and its follow-up studies
advanced our understanding of emergent phenomena based on solvable models in $D>1$.
It was believed that spin rotational symmetry is essential for QSL, and a spin system will tend to be ordered if the spin rotational symmetry is broken.
As a counterexample, Kitaev demonstrated an exactly solvable spin model with strong spin-orbit coupling\cite{Kitaev,Jackeli},
which destroys the spin rotational symmetry, can host fractionalized spin excitations - deconfined spinons on top of quantum paramagnetic ground states.
After that, great theoretical efforts have been devoted to searching for its exactly solvable generalizations,
including various two dimensional ($2D$)\cite{yao2007,yang,baskaran2009,tikhonov2010} and three dimensional ($3D$)\cite{si,mandal,ryu,Trebst15} lattices, $SU(2)$ invariant systems\cite{wang,yao2011,lai2011},
multiple-spin interactions\cite{lee2007,yu2008} and higher spin systems\cite{yao2009,wu,chern,chua,nakai,nussinov2009,nussinov2013}.
However, except a spin-$3/2$ model on square lattice proposed by Yao et. al. \cite{yao2009},
all these exactly solvable models are not Mott insulators in the strict sense, because they carry an even number of half spins per unit cell, namely, the total spin quanta in a unit cell is an integer rather than a half-integer.
Thus, a spin-$1/2$ exactly solvable Mott insulator model, which carries an odd number of half spins per unit cell, is still highly desirable.

In this paper, we proposed an exactly solvable quantum spin-$1/2$ model on a $2D$ lattice, which consists of three sites in each unit cell and hosts algebraic QSL ground states with point nodal spinon excitations.
The paper is organized as follows. We present the model Hamiltonian and analytical methods in Section~\ref{sec:model},
and main results in Section~\ref{sec:result}.
Section ~\ref{sec:conclusion} is devoted to discussion and conclusions.
We also provide appendices for detailed calculations.

\section{Model and Methods}\label{sec:model}

\begin{figure}[hptb]
\begin{center}
\includegraphics[width=8.2cm]{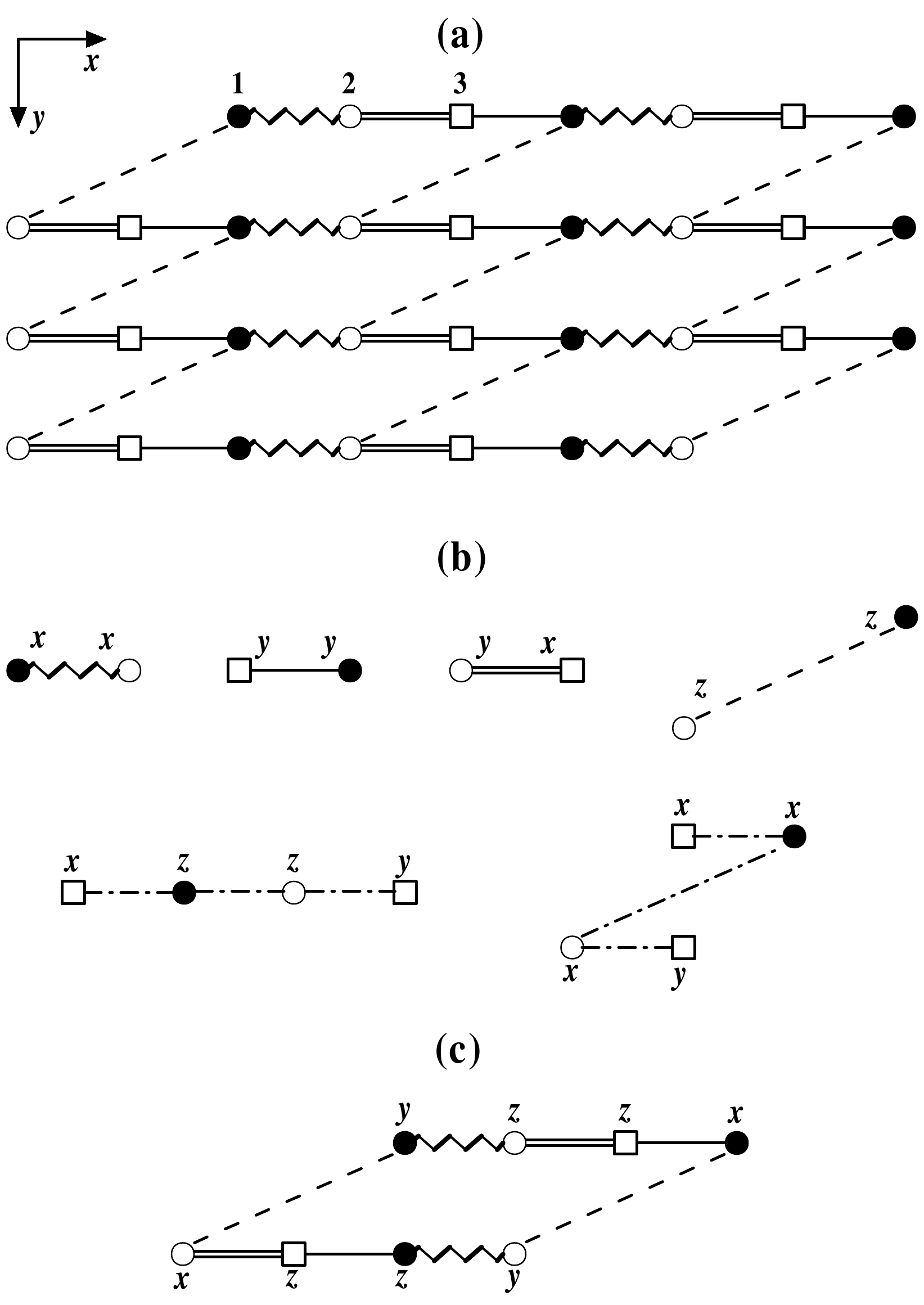}
\end{center}
\caption{(a) A inclined brick-wall lattice with three sites in each unit cell. A site $n$ is labeled by the unit cell $\vec{r}=(n_x,n_y)$ and the sublattice index $\mu_n=1,2,3$.
Black circles, white circles and white squares represent sublattices $1$, $2$ and $3$ respectively.
(b) Wiggle lines denote $\sigma^{x}_{n}\sigma^{x}_{m}$ terms, solid lines denote $\sigma^{y}_{n}\sigma^{y}_{m}$ terms,
double-solid lines denote $\sigma^{y}_{n}\sigma^{x}_{m}$ terms, and dashed lines denote $\sigma^{z}_{n}\sigma^{z}_{m}$ terms.
The $t_x$ and $t_y$ terms in Eq.\eqref{eq:H1} are defined along the horizontal chain and the zigzag chain respectively.
(c) A plaquette where the loop operator $\hat{\phi}_p$ in Eq.\eqref{eq:Wp} is defined.
}
\label{fig:latt}
\end{figure}

{\em Model Hamiltonian.} Consider a $L_x \times L_y \times 3$ inclined brick-wall lattice as plotted in FIG. \ref{fig:latt}, where each unit cell consists of three sites.
Labeling a site $n$ by the unit cell $\vec{r}_{n}=n_{x}\hat{x}+n_{y}\hat{y}$ and the sublattice index $\mu_n=1,2,3$, the Mott insulator model Hamiltonian is given by
\begin{subequations}\label{eq:H}
\begin{eqnarray}
   H  &=& H_{0}+H_{1}, \\
H_{0} &=& \sum_{\vec{r}}J_{x}\sigma_{\vec{r},1}^{x}\sigma_{\vec{r} ,2}^{x}+J_{yx}\sigma_{\vec{r},2}^{y}\sigma_{\vec{r},3}^{x}+J_{y}\sigma_{\vec{r},3}^{y}\sigma_{\vec{r}+\hat{x},1}^{y}\nonumber\\
      & & + J_{z}\sigma _{\vec{r},1}^{z}\sigma_{\vec{r}-\hat{x}+\hat{y},2}^{z}, \label{eq:H0} \\
H_{1} &=&\sum_{\vec{r}}t_{x} \sigma_{\vec{r},3}^{x}\sigma_{\vec{r}+\hat{x},1}^{z}\sigma_{\vec{r}+\hat{x},2}^{z}\sigma_{\vec{r}+\hat{x},3}^{y} \nonumber \\
      & & + t_{y} \sigma_{\vec{r},3}^{x}\sigma_{\vec{r}+\hat{x},1}^{x}\sigma_{\vec{r}+\hat{y},2}^{x}\sigma_{\vec{r}+\hat{y},3}^{y}, \label{eq:H1}
\end{eqnarray}
\end{subequations}
where $\sigma_{\vec{r},\mu_n}^{\alpha}$ is the Pauli matrix at site $n$ with $\alpha=x,y,z$.
$H_0$ consists of only two-spin interactions and $H_1$ consists of only four-spin interactions.
The values of coupling constants $J_x$, $J_y$, $J_{yx}$, $J_z$, $t_x$ and $t_y$ can be chosen as any real number.
Note that there are even number of spin operators in each term such that the time reversal symmetry (TRS) is guaranteed.

{\em Loop operator in a plaquette.} Similar to Kitaev honeycomb model, a loop operator $\hat{\phi}_{p}$ can be defined in a plaquette $p$ (as shown in FIG. \ref{fig:latt}(c)) as follows,
\begin{equation}
\hat{\phi}_{p}=-\sigma^{y}_{\vec{r},1}\sigma^{z}_{\vec{r},2}\sigma^{z}_{\vec{r},3}\sigma^{x}_{\vec{r}+\hat{x},1}\sigma^{y}_{\vec{r}+\hat{y},2}\sigma^{z}_{\vec{r}+\hat{y},1}\sigma^{z}_{\vec{r}-\hat{x}+\hat{y},3}\sigma^{x}_{\vec{r}-\hat{x}+\hat{y},2}. \label{eq:Wp}
\end{equation}
Note that $\hat{\phi}_{p}$ commutes with $H$ and $\hat{\phi}_p^2=1$, whose eigenvalue $\phi_p=\pm 1$ is a good quantum number. As it can be seen below, $\phi_p$ serves as a $Z_2$ flux. 

{\em Exact solvability.} The spin Hamiltonian in Eq. \eqref{eq:H} can be exactly solved with the help of the Jordan-Wigner transformation\cite{Jordan-Wigner}, 
which was applied to solve original Kitaev honeycomb model\cite{Feng,Chen,Chen-Nussinov} in addition to the exact solution through four Majorana decomposition\cite{Kitaev}.
This elegant method enables a fermionization of the spin model without redundant degrees of freedom. 
To implement this transformation, we sort all the sites as follows: 
for two sites $m$ and $n$, (1) if $m_y<n_y$ then $m<n$; (2) if $m_y=n_y$ and $m_x<n_x$ then $m<n$; (3) if $m_x=n_x, m_y=n_y$ and $\mu_m<\mu_n$, then $m<n$.
By this definition of site order, the Jordan-Wigner transformation can be adapted,
\begin{subequations}
\begin{eqnarray}
\sigma_{m}^{+} & = & c_{m}^{\dagger}e^{i\pi\sum_{l<m}\hat{n}_{l}},\\
\sigma_{m}^{z} & =& 2\hat{n}_{m}-1,
\end{eqnarray}
\end{subequations}
where $\sigma_{m}^{+}=\frac{1}{2}(\sigma_{m}^{x}+i\sigma_{m}^{y})$ is the spin raising operator, $c_{m}^{\dagger}$ is the creation operator for the spinless fermion at site $m$, 
and $\hat{n}_{m}=c_{m}^{\dagger}c_{m}$ is the fermion occupation number. 
Then we decompose the complex fermion $c_n$ into two Majorana fermions (MFs) $\eta_{n}$ and $\beta_{n}$ as follows: (1) for $\mu_n=1$, $\eta_{n}=c_{n}^{\dagger}+c_{n}$ 
and $\beta_{n}=i\left(c_{n}^{\dagger}-c_{n}\right)$, (2) for $\mu_n=2\, \text{or}\, 3$, $\eta_{n}=i\left(c_{n}^{\dagger}-c_{n}\right)$ and $\beta_{n}=c_{n}^{\dagger}+c_{n}$.
After the Jordan-Wigner transformation, the two-spin terms in Eq. \eqref{eq:H0} become
\begin{subequations}\label{eq:JW1}
\begin{eqnarray}
\sigma_{\vec{r},1}^{x}\sigma_{\vec{r},2}^{x}          & = & -i\beta_{\vec{r},1}\beta_{\vec{r},2}, \label{eq:JWxx}\\
\sigma_{\vec{r},2}^{y}\sigma_{\vec{r},3}^{x}          & = & -i\beta_{\vec{r},2}\beta_{\vec{r},3}, \label{eq:JWyx}\\
\sigma_{\vec{r},3}^{y}\sigma_{\vec{r}+\hat{x},1}^{y}  & = & i\beta_{\vec{r},3}\beta_{\vec{r}+\hat{x},1}, \label{eq:JWyy}\\
\sigma_{\vec{r},1}^{z}\sigma_{\vec{r}-\hat{x}+\hat{y},2}^{z}      & = &i\hat{D}_{\vec{r}}\beta_{\vec{r},1}\beta_{\vec{r}-\hat{x}+\hat{y},2}, \label{eq:JWzz}
\end{eqnarray}
the four-spin terms in Eq. \eqref{eq:H1} become
\begin{eqnarray}
\sigma_{\vec{r},3}^{x}\sigma_{\vec{r}+\hat{x},1}^{z}\sigma_{\vec{r}+\hat{x},2}^{z}\sigma_{\vec{r}+\hat{x},3}^{y} & = & i\eta_{\vec{r},3}\eta_{\vec{r}+\hat{x},3}, \label{eq:JWTx} \\
\sigma_{\vec{r},3}^{x}\sigma_{\vec{r}+\hat{x},1}^{x}\sigma_{\vec{r}+\hat{y},2}^{x}\sigma_{\vec{r}+\hat{y},3}^{y} & = & -i\hat{D}_{\vec{r}+\hat{x}}\eta_{\vec{r},3}\eta_{\vec{r}+\hat{y},3}, \label{eq:JWTy}
\end{eqnarray}
and the loop operator $\hat{\phi}_{p}$ becomes
\begin{equation} 
\hat{\phi}_{p}= \hat{D}_{\vec{r}} \hat{D}_{\vec{r}+\hat{x}}, \label{eq:JWWp}
\end{equation}
\end{subequations}
where $\hat{D}_{\vec{r}}= i\eta_{\vec{r},1}\eta_{\vec{r}-\hat{x}+\hat{y},2}$. 
It is easy to verify that $\hat{D}_{\vec{r}}$ commute with each other and with the Hamiltonian, and $\hat{D}_{\vec{r}}^{2}=1$.
So that we can replace the operator $\hat{D}_{\vec{r}}$ by its eigenvalues $D_{{\vec{r}}}=\pm1$, which can be viewed as a $Z_2$ background field. Thus $\phi_p$ is the corresponding $Z_2$ flux in a plaquette.
The eigenstates of the Hamiltonian can be divided into different sectors of total Hilbert space according to the sets of eigenvalues $\left\{ D_{\vec{r}}\right\}$. 
In each sector, all the two-spin and four-spin terms in Eq. \eqref{eq:H} are transformed to quadratic MF terms and the total Hamiltonian is exactly diagonalizable.

{\em Lift local degeneracy.} For simplicity, we only consider situations when both $L_x$ and $L_y$ are even numbers hereafter. 
When $t_x=t_y=0$, there exist $2^{L_x L_y/2}$-fold degeneracy for $H_0$ due to the local symmetries generated by $\eta_{\vec{r},3}$ operators.
This local degeneracy will be lifted by nonvanishing $H_1$ without breaking translational symmetry explicitly.
It is easy to see from Eqs. \eqref{eq:H1}, \eqref{eq:JWTx} and \eqref{eq:JWTy} that $H_1$ describes free MFs $\eta_{\vec{r},3}$ coupled to the $Z_2$ background fields $D_{\vec{r}}$ on a square lattice.
Numerically we find that the ground states of the system are always $\pi$-flux states, i.e., ${\phi_{p}=-1}$ everywhere.
The energy dispersion of such a $\pi$-flux ground state reads,
\begin{equation}
\epsilon_{\eta_{3}}(\vec{k})=\pm\sqrt{(t_x \sin k_x)^2 + (t_y \sin k_y)^2}, \label{eq:Eketa3pi}
\end{equation}
which has Dirac nodes.
Thus the enormous local degeneracy is lifted and possible residual degeneracy for ground states is topological degeneracy as we will discuss later.

{\em Periodic boundary condition.} Note that Eqs. \eqref{eq:JWyy}, \eqref{eq:JWTx}, \eqref{eq:JWTy} and \eqref{eq:JWWp} are valid for open boundary condition (OBC) only. 
Under periodic boundary condition (PBC), additional boundary terms will appear as
\begin{subequations}\label{eq:JW2}
\begin{equation}
\sigma_{L_x,n_y,3}^{y}\sigma_{1,n_y,1}^{y}  =  i\beta_{L_x,n_y,3}\beta_{1,n_y,1}\hat{F}_{n_y}, 
\end{equation}
and 
\begin{eqnarray}
&  &\sigma_{L_{x},n_{y},3}^{x}\sigma_{1,n_{y},1}^{z}\sigma_{1,n_{y},2}^{z}\sigma_{1,n_{y},3}^{y} \nonumber\\
& = & i\eta_{L_{x},n_{y},3}\eta_{1,n_{y},3}\hat{F}_{n_{y}},
\end{eqnarray}
and
\begin{eqnarray}
&  &\sigma_{L_{x},n_{y},3}^{x}\sigma_{1,n_{y},1}^{x}\sigma_{L_{x},n_{y}+1,2}^{x}\sigma_{L_{x},n_{y}+1,3}^{y} \nonumber \\
&= &-i\hat{D}_{1,n_{y}}\eta_{L_x,ny,3}\eta_{L_x,n_y+1,3}\hat{F}_{n_{y}},
\end{eqnarray}
and the flux operators on the edge plaquettes become
\begin{align} 
\hat{\phi}_{p} & = \hat{D}_{L_x,n_y} \hat{D}_{1,n_y} \hat{F}_{n_y},\label{eq:JWWp2} \\
\hat{\phi}_{p} & = \hat{D}_{1,n_y} \hat{D}_{2,n_y} \hat{F}_{n_y+1},\label{eq:JWWp3}
\end{align}
\end{subequations}
where $\hat{F}_{n_y}=e^{i\pi \hat{N}_{n_y}}$ and $\hat{N}_{n_y}=\sum_{n_x,\mu}\hat{n}_{n_x,n_y,\mu}$ are fermion parity and occupation number in the $n_y$-th row respectively. 
As pointed out by Yao et al.\cite{yao2007}, $\hat{D}_{\vec{r}}$ does not commute with $\hat{F}_{n_y}$ thereby $H$ under PBC.
Nevertheless, we can choose $\left\{\phi_p\right\}$ as the good quantum numbers instead of $\left\{D_{\vec{r}}\right\}$.
In addition, there exists two extra degrees of freedom given by the global fluxes, $\hat{\Phi}_{x}= \hat{F}_{n_y=1}$ along the $x$-direction 
and $\hat{\Phi}_{y}=\prod_{n_y}\hat{D}_{\vec{r}=(1,n_y)}$ along the $y$-direction\cite{globalPhi}. It is easy to see that $\hat{\Phi}_{x}^2=\hat{\Phi}_{y}^2=1$ and   
corresponding eigenvalues read $\Phi_{x}=\pm 1$ and $\Phi_{y}=\pm 1$.
Since both $\hat{\Phi}_{x}$ and $\hat{\Phi}_{x}$ commute with $\hat{\phi}_{p}$ and $H$, 
we can divide the total Hilbert space into subspaces according to the sets of good quantum numbers $\{\phi_{p},\Phi_{x},\Phi_{y}\}$.
In each subspace, the spin Hamiltonian $H$ is transformed to quadratic MF terms.

{\em Degrees of freedom and physical spin states.} Now let us consider the degrees of freedom in a $L_x\times L_y \times 3$ system. There are total $2^{3L_x L_y}$ possible spin states. 
All the sets of $\{\phi_{p},\Phi_x,\Phi_y\}$ give rise to $2^{L_x L_y+1}$ degrees of freedom, subject to the constraint $\prod_{p}\hat{\phi}_p=1$, where the product is over all the plaquettes. 
For a given set of $\{\phi_{p},\Phi_x,\Phi_y\}$, solving the Hamiltonian consisting of MFs $\{\eta_{\vec{r},3},\beta_{\vec{r},1},\beta_{\vec{r},2},\beta_{\vec{r},3}\}$ results in $2^{2L_x L_y}$ eigenstates in each subspace.
Hence there are total $2^{3L_x L_y+1}$ states in the fermion representation, which doubles the number of physical spin states.
It means that half of the states in the fermion representation are unphysical indeed. These unphysical states originate as follows. 
The total fermion number parity $\hat{F}=\prod_{n_y}\hat{F}_{n_y}$ anticommutes with all the MFs.
When one solves the Hamiltonian with a given set of $\{\phi_{p},\Phi_x,\Phi_y\}$, the the eigenvalue of $\hat{F}$, $F$, is presumed.
But the MFs $\eta_{\vec{r},2}$ and $\beta_{\vec{r},\mu}$ will change the sign of $F$, resulting in half unphysical states with incompatible $F$ value in each sector.
The unphysical states can be removed by the projection $\hat{P}=(1+F\hat{F})/2$\cite{Projection}. It means that {\it a physical spin excitation should be composed of even number of fermions}. 
We have examined this in a $2\times 2\times 3$ system by exact diagonalization.

\section{Results}\label{sec:result}

{\em Ground states and topological degeneracy.}
It turns out that there are four {\em unprojected} fermion degenerate ground states on a torus characterized by the global fluxes $\Phi_x=\pm 1$ and $\Phi_y=\pm 1$, 
as we have searched numerically for all the possible values of $J_x$, $J_{xy}$, $J_y$ and $J_z$. 
These four fermion ground states are all $\pi$-flux states with $\phi_p=-1$ on every plaquette. 
However, the ground state with $\Phi_x=\Phi_y=1$ is eliminated by the projection $\hat{P}$, and the other three will survive and give rise to physical spin states\cite{DGS}.
For a finite system, the energy difference between these ground states, $\Delta$, are finite, and $\Delta\propto 1/L$, where $L$ is the linear size of the lattice.
The degeneracy is robust against any local perturbations, as we have examined by adding disordered $J_x$, $J_{xy}$, $J_y$ and $J_z$ terms to $H_0$ in Eq. \eqref{eq:H0}, 
which will not spoil the exact solvability. Thus, this degeneracy is a {\em three-fold topological degeneracy}.

{\em Bulk excitations.} The bulk excitations on top of the $\pi$-flux ground states are all gapless. 
As mentioned, $H_1$ in Eq. \eqref{eq:H1} describes free MFs on a square lattice (for $\eta_{\vec{r},3}$) and gives rise to Dirac spectra in Eq.\eqref{eq:Eketa3pi}.
Now we focus on the gapless excitations from $\beta_{\vec{r},\mu}$ MFs. 
The uniform $\pi$-fluxes give rise to magnetic unit cells which double the primitive cells, and reduce the Brillouin zone by a factor 2, although the translational symmetry does not break physically.
Denote the two primitive cells in a magnetic unit cell as $A$ and $B$, we can define a 6-component spinor field $\Gamma_{\vec{q}} = \left(\beta^{A}_{\vec{q},1}, \beta^{A}_{\vec{q},2}, \beta^{A}_{\vec{q},3}, \beta^{B}_{\vec{q},1}, \beta^{B}_{\vec{q},2}, \beta^{B}_{\vec{q},3}\right)$, 
which is the Fourier transform of the six $\beta$-MFs in a magnetic unit cell. 
Note that $\beta_{\vec{q},\mu}^{A,B}$ satify anticommutation relations $\left\{ \beta_{q,\mu'}^{A/B}\beta_{q',\mu'}^{A/B\dagger}\right\} =\delta_{q,q'}\delta_{\mu,\mu'}\delta_{A,B}$, and the condition $\beta_{-\vec{q},\mu}^{A,B}=\beta_{\vec{q},\mu}^{A,B\dagger}$ is imposed by the self-hermitiancity of MFs. 
The $\pi$-flux Hamiltonian for $\beta$-MFs reads $H_{\pi-F} = \frac{i}{2}\sum_{\vec{q}}\Gamma_{\vec{q}}^{\dagger}h(\vec{q})\Gamma_{\vec{q}}$, where $h(q)$ is a $6\times 6$ matrix given in the following,
\begin{widetext}
\begin{equation}\label{eq:hq}
h(\vec{q}) = \left(\begin{array}{cccccc}
0 & -J_{x} & 0 & 0 & J_{z}e^{i(q_{y}-q_{x})} & -J_{y}e^{-iq_{x}} \\
J_{x} & 0 & -J_{yx} & J_{z}e^{-iq_{y}} & 0 & 0 \\
0 & J_{yx} & 0 & J_{y} & 0 & 0 \\
0 & -J_{z}e^{iq_{y}} & -J_{y} & 0 & -J_{x} & 0 \\
-J_{z}e^{-i(q_{y}-q_{x})} & 0 & 0 & J_{x} & 0 & -J_{yx} \\
J_{y}e^{iq_{x}} & 0 & 0 & 0 & J_{yx} & 0 \\
\end{array}\right).
\end{equation}
\end{widetext}

\begin{figure}[hptb]
\begin{center}
\includegraphics[width=8cm]{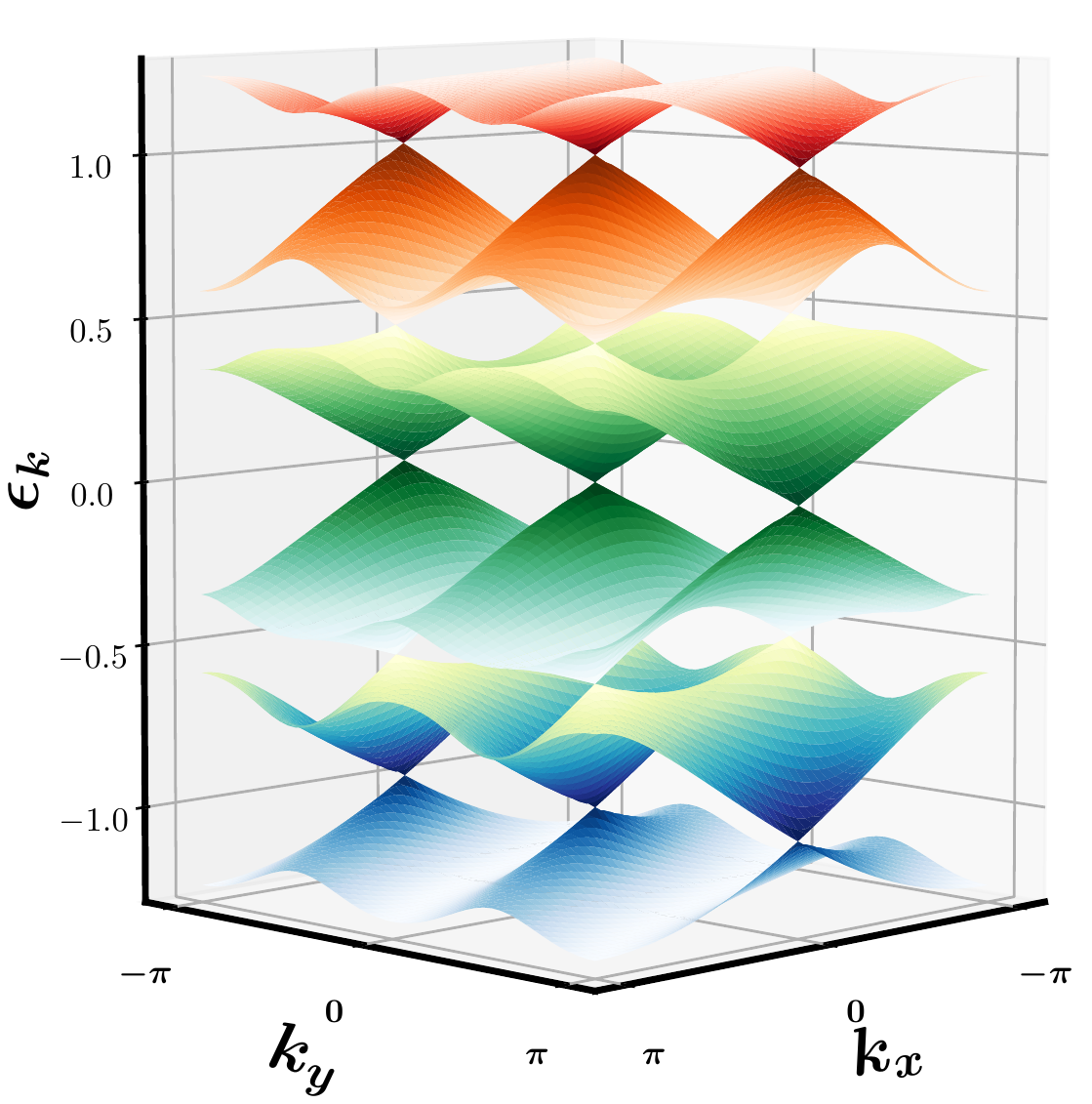}
\end{center}
\caption{(Color online) Six bands from $\beta$ Majorana fermions for the $\pi$-flux state in the reduced Brillouin zone, where the coupling parameters are chosen as $J_x=J_{xy}=J_{y}=J_{z}=1$.}
\label{fig:band}
\end{figure}

These six $\beta$-MFs form six bands in the reduced Brillouin zone as illustrated in FIG. \ref{fig:band}. 
The three upper bands with positive energy are quasi-particle bands and the three lower bands with negative energy are quasi-hole bands.
Each band connects to another band at two nodal points $(k_x,k_y)=(0,0)$ and $(0,\pi)$ and is separated from other bands. The energy dispersions around these nodal points are Dirac-like.
All these features do not depend on which degenerate ground state is studied and how the parameters $J_x$, $J_{xy}$, $J_y$ and $J_z$ are chosen. 
The ground states are obtained by filling all the quasi-hole states. And the bulk excitations are gapless with two nodal points.

\begin{figure}[hptb]
\begin{center}
\includegraphics[width=8cm]{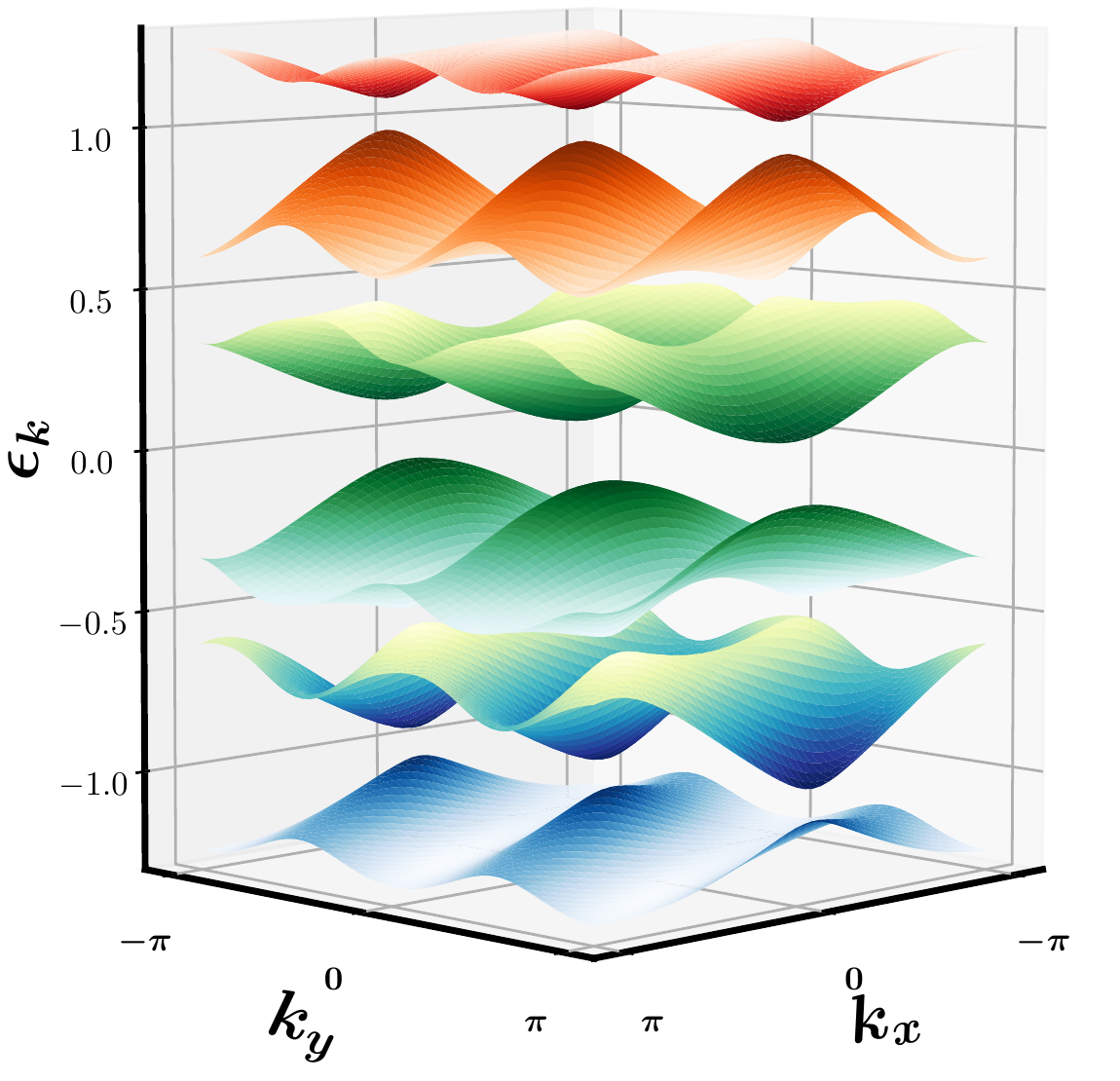}
\end{center}
\caption{(Color online) A magnetic field $\vec{h}=(h,h,h)$ will open bulk gaps and separate the six $\beta$-MF bands from each other. The ground state is a $\pi$-flux state with the coupling parameters $J_x=J_{xy}=J_{y}=J_{z}=1$, and $h^{3}/\Delta_{v}^2=0.1$.}
\label{fig:band2}
\end{figure}

{\em Breaking the TRS.} An external magnetic field will break the TRS and open a bulk energy gap,
which allows us to evaluate Chern numbers and study possible in-gap states in the presence of vortices. 
It is similar to Kitaev honeycomb model for which the Zeeman term $\sum_{n}\vec{h}\cdot\vec{\sigma}_n$ spoils the exact solvability but some leading order perturbation terms do not\cite{Kitaev,lee2007},
where $\vec{h}=(h_x,h_y,h_x)$ is the magnetic field.
We consider the following third order perturbations in the vortex-free subspace which remains the exact solvability,
\begin{eqnarray}
H^{\prime} & = & \frac{h^{3}}{\Delta_{v}^2}\sum_{\vec{r}} \sigma^{x}_{\vec{r},1}\sigma^{z}_{\vec{r},2}\sigma^{x}_{\vec{r},3}+\sigma^{y}_{\vec{r},2}\sigma^{z}_{\vec{r},3}\sigma^{y}_{\vec{r}+\hat{x},1} \nonumber\\
&&+\sigma^{y}_{\vec{r},3}\sigma^{z}_{\vec{r}+\hat{x},1}\sigma^{x}_{\vec{r}+\hat{x},2}+\sigma^{z}_{\vec{r},1}\sigma^{x}_{\vec{r}-\hat{x}+\hat{y},1}\sigma^{y}_{\vec{r}-\hat{x}+\hat{y},2} \nonumber\\
&&+\sigma^{y}_{\vec{r},1}\sigma^{x}_{\vec{r},2}\sigma^{z}_{\vec{r}-\hat{x}+\hat{y},2}+\sigma^{y}_{\vec{r},3}\sigma^{x}_{\vec{r}+\hat{x},1}\sigma^{z}_{\vec{r}+\hat{y},2} \nonumber\\
&&+\sigma^{z}_{\vec{r},1}\sigma^{x}_{\vec{r}-\hat{x}+\hat{y},2}\sigma^{x}_{\vec{r}-\hat{x}+\hat{y},3}, \label{eq:H3}
\end{eqnarray}
where we have set $h_x=h_y=h_z=h$ for simplicity and $\Delta_{v}$ is the vortex excitation gap.
$H^{\prime}$ commute with $\hat{\phi}_p$, $\hat{\Phi}_{x}$, $\hat{\Phi}_{y}$ and $H$, and can be transformed to quadratic $\beta$-MF terms.
For a small perturbation, e.g., $h^{3}/\Delta_{v}^2=0.1$, the ground states are still $\pi$-flux states and of three-fold topological degeneracy.
$H^{\prime}$ will open bulk gaps in $\beta$-MF bands and separate the six bands from each other in energy as demonstrated in FIG.~\ref{fig:band2}. 
Thus Chern numbers can be evaluated for each fermion band\cite{CN05}.  
Using $C\#$ to denote a sequence of Chern numbers from the lowest band to the highest band, it follows that
\begin{equation}
C\#=(-1,-1,1,-1,1,1). \label{eq:CN}
\end{equation}
Note that the third order perturbation will not affect the $\eta_3$-MF band in $H_{1}$. To open a bulk gap in the $\eta_3$-MF band, the fifth order perturbation is required.

\begin{figure}[hptb]
\begin{center}
\includegraphics[width=8.4cm]{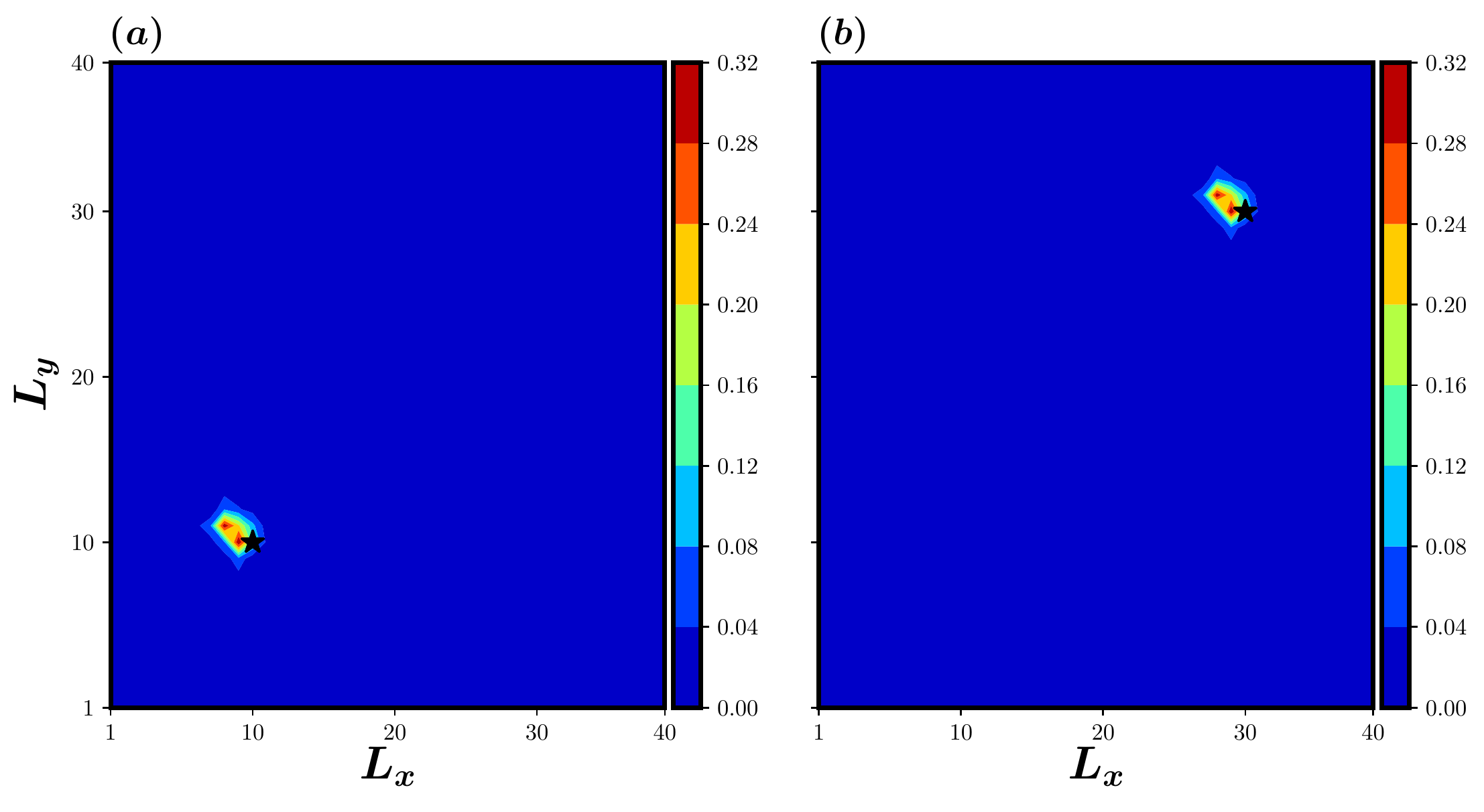}
\end{center}
\caption{(Color online) The density distribution of the MZMs. Two vortices are created at two black-star plaquettes. Two MZMs are localizd at (a) left-down and (b) right-up vortex core respectively.}
\label{fig:MZM}
\end{figure}

{\em Vortices.} 
The nontrivial topology motivates us to study the vortex excitations on top of the gapped ground states.
A vortex can be created by setting $\phi_p=1$ in one plaquette and remaining $\phi_p=-1$ in all the others.
Note that the creation of odd number of vortices can be achieved only under OBC. 
For PBC, only even number of vortices can be created as the product of all the loop operaors should be identity.
We study a pair of well separated votrices, and find two Majorana zero modes (MZMs) localized at the two vortex core centers as shown in FIG.~\ref{fig:MZM}.
Due to the presence of MZMs, it seems that there is extra double topological degeneracy for given global fluxes $\Phi_x$ and $\Phi_y$.
However the projection $\hat{P}$ will remove one of them as MZMs will change the fermion number parity $F$.
Thus the true ground states degeneracy on a $2D$ torus is $3$-fold, which is consistent with the non-Abelian Ising topological order nature of the gapped phase.
Regarding global fluxes and the projection, the ground state degenercay will be $2^{n+1}$-fold in the presence of $2n$ well separated vortices,
which indicates non-Abelian statistics of the vortices and is consistent with the odd total Chern number in filled bands.
The growing degeneracy reflects the quantum dimension of MZMs is $\sqrt{2}$ as non-Abelian anyons in Kitaev honeycomb model\cite{Kitaev}, as well as in triangular-honeycomb model proposed by Yao et al.\cite{yao2007}.

\section{Conclusion and Discussions}\label{sec:conclusion}

In summary, we proposed an exactly solvable spin-1/2 model in $2D$, which is a Mott insulator model in the strict sense and hosts algebraic QSL ground states. 
The ground states are three-fold topological degeneracy on a torus. The bulk spinon excitations consist of two Dirac nodes at $(0,0)$ and $(0,\pi)$.
Breaking the TRS by an external magnetic field will open a spinon gap, and $Z_2$ vortices will carry non-Abelian braiding statistics.
It is worth noting that the $\sigma^y\sigma^x$ terms are crucial to construct the exactly solvable model with odd number of half-spins per unit cell. These off-diagonal terms also make the models beyond the category of compass models\cite{nussinov2013} and can be used to construct further exactly solvable models\cite{Miao}.
Finally, we would like to point out that the spin-1/2 anisotropic interaction is easier to be realized than those of higher spins, and corresponding lattice may be found in some metal-organic frameworks (also known as coordination polymers)\cite{MOF}.

\emph{Acknowledgement.}
We thank Long Zhang, Meng Cheng and especially Hong Yao for helpful discussions, and D. H. Xu for his help in Chern number calculation.
JJM is supported by Postdoctoral Science Foundation of China (No.119103S284).
FW acknowledges support from National Key Basic Research Program of China (No.2014CB920902) and
National Key Research and Development Program of China (No.2017YFA0302904).
YZ is supported by National Key Research and Development Program of China (No.2016YFA0300202),
National Natural Science Foundation of China (No.11774306), 
and the Strategic Priority Research Program of Chinese Academy of Sciences (No. XDB28000000), 
and the Fundamental Research Funds for the Central Universities in China.
FCZ is supported by NSFC grant 11674278 and the CAS Center for Excellence in Topological Quantum Computation.


\appendix

\section{Breaking the time reversal symmetry: Kitaev honeycomb model}
For the Kitaev honeycomb model, Kitaev introduced the following magnetic field
\begin{equation}
V=\sum_{j}\vec{h}\cdot\vec{\sigma_j}
\end{equation}
Kitaev want to obtain the effective Hamiltonian in the vortex free sector. The first order perturbation vanishes
\begin{equation}
H_{eff}^{\left(1\right)}=\varPi_{0}V\varPi_{0}
\end{equation}
as the perturbation $V$ will introduce the vortex.
The second order perturbation does not vanish 
\begin{equation}
H_{eff}^{\left(2\right)}=\varPi_{0}VG'_{0}\left(E_{0}\right)V\varPi_{0}
\end{equation}
as one perturbation $V$ can create the vortex and another vortex can annihilate the vortex correspondingly and then the system is still vortex free.  However second order perturbation can not open the gap as it preserves the time-reversal symmetry.
Thus the leading order perturbation opening the gap is the third order perturbation 
\begin{align}
H_{eff}^{\left(3\right)} & =\varPi_{0}VG'_{0}\left(E_{0}\right)VG'_{0}\left(E_{0}\right)V\varPi_{0}\nonumber \\
 & =\varPi_{0}\frac{V^{3}}{\Delta_{v}^{2}}\varPi_{0}.
\end{align}
The leading effective Hamiltonian becomes
\begin{equation}
H_{eff}^{\left(3\right)}=\frac{h_{x}h_{y}h_{z}}{\Delta_{v}^{2}}\sum_{\left\langle j,k,l\right\rangle }\sigma_{j}^{x}\sigma_{k}^{y}\sigma_{l}^{z}
\end{equation}
However to project onto the vortex free sector, the summation can be only performed over spin triples. There are two kinds of triples, one is 
\begin{equation}
\left(\sigma_{j}^{x}\sigma_{l}^{x}\right)\left(\sigma_{l}^{y}\sigma_{k}^{y}\right)\sim\sigma_{j}^{x}\sigma_{l}^{z}\sigma_{k}^{y}
\end{equation}
another is 
\begin{equation}
\left(\sigma_{m}^{x}\sigma_{j}^{x}\right)\left(\sigma_{k}^{y}\sigma_{m}^{y}\right)\left(\sigma_{m}^{z}\sigma_{l}^{z}\right)\sim\sigma_{j}^{x}\sigma_{k}^{y}\sigma_{l}^{z}.
\end{equation}
Kitaev notes the second kind of triples correspond to four-fermion terms and therefore do not directly influence the spectrum.

\section{Breaking the time reversal symmetry: Mott insulator model}

We can apply magnetic field $\vec{h}$ to the system in the form of $\sum_{\vec{r}}\vec{h}\cdot\vec{\sigma}_{\vec{r}}$. For simplicity, we assume that $\vec{h}=(h,h,h)$. By the perturbation theory introduced by Kitaev, we obtain the effective Hamiltonian as 
\begin{equation}
\begin{split}
H_{eff} = \frac{h^{3}}{\Delta_{v}^2}\sum_{\vec{r}}&\sigma^{x}_{\vec{r},1}\sigma^{z}_{\vec{r},2}\sigma^{x}_{\vec{r},3}+\sigma^{y}_{\vec{r},2}\sigma^{z}_{\vec{r},3}\sigma^{y}_{\vec{r}+\hat{x},1}\\
+&\sigma^{y}_{\vec{r},3}\sigma^{z}_{\vec{r}+\hat{x},1}\sigma^{x}_{\vec{r}+\hat{x},2}+\sigma^{z}_{\vec{r},1}\sigma^{x}_{\vec{r}-\hat{x}+\hat{y},1}\sigma^{y}_{\vec{r}+\hat{x}+\hat{y},2}\\
+&\sigma^{y}_{\vec{r},1}\sigma^{x}_{\vec{r},2}\sigma^{z}_{\vec{r}-\hat{x}+\hat{y},2}+\sigma^{y}_{\vec{r},3}\sigma^{x}_{\vec{r}+\hat{x},1}\sigma^{z}_{\vec{r}+\hat{y},2}\\
+&\sigma^{z}_{\vec{r},1}\sigma^{x}_{\vec{r}-\hat{x}+\hat{y},2}\sigma^{x}_{\vec{r}-\hat{x}+\hat{y},3}.\\
\end{split}
\end{equation}
Note to obtain the nontrivial Hamiltonian, we should at leat consider the third order perturbations.
The Hamiltonian $H_{eff}$ contains three spin interactions and break the TRS explicitly.
In the same spirit of Kitaev, we only retain the terms which correspond to quadratic $\beta$ MFs terms. By fermionization, we obtain
\begin{subequations}
	\begin{eqnarray}
	\sigma^{x}_{\vec{r},1}\sigma^{z}_{\vec{r},2}\sigma^{x}_{\vec{r},3} & = & i\beta_{\vec{r},1}\beta_{\vec{r},3},\\
	\sigma^{y}_{\vec{r},2}\sigma^{z}_{\vec{r},3}\sigma^{y}_{\vec{r}+\hat{x},1}  & = & -i\beta_{\vec{r},2}\beta_{\vec{r}+\hat{x},1}, \\
	\sigma^{y}_{\vec{r},3}\sigma^{z}_{\vec{r}+\hat{x},1}\sigma^{x}_{\vec{r}+\hat{x},2} & = & i\beta_{\vec{r},3}\beta_{\vec{r}+\hat{x},2}, \\
	\sigma^{z}_{\vec{r},1}\sigma^{x}_{\vec{r}-\hat{x}+\hat{y},1}\sigma^{y}_{\vec{r}-\hat{x}+\hat{y},2}     & = &i\hat{D}_{\vec{r}}\beta_{\vec{r},1}\beta_{\vec{r}-\hat{x}+\hat{y},1},\\
	\sigma^{y}_{\vec{r},1}\sigma^{x}_{\vec{r},2}\sigma^{z}_{\vec{r}-\hat{x}+\hat{y},2} & = &-i\hat{D}_{\vec{r}}\beta_{\vec{r},2}\beta_{\vec{r}-\hat{x}+\hat{y},2},\\
	\sigma^{y}_{\vec{r},3}\sigma^{x}_{\vec{r}+\hat{x},1}\sigma^{z}_{\vec{r}+\hat{y},2} & = & i\hat{D}_{\vec{r}+\hat{x}}\hat{F}_{r_{y}}\beta_{\vec{r},3}\beta_{\vec{r}+\hat{y},2},\\
	\sigma^{z}_{\vec{r},1}\sigma^{x}_{\vec{r}-\hat{x}+\hat{y},2}\sigma^{x}_{\vec{r}-\hat{x}+\hat{y},3} & = &
	i\hat{D}_{\vec{r}}\beta_{\vec{r},1}\beta_{\vec{r}-\hat{x}+\hat{y},3}.
	\end{eqnarray}
\end{subequations}
In momentum space, the perturbative Hamiltonian can be rewritten as 
\begin{widetext}
\begin{equation}
h_{eff}(\vec{q}) = \frac{h^{3}}{\Delta_{v}^{2}}\left(\begin{array}{cccccc}
0 & 0 & 1 & e^{i(q_{y}-q_{x})}+e^{-iq_{y}} & e^{-iq_{x}} & e^{i(q_{y}-q_{x})} \\
 & 0 & e^{-iq_{y}} & -1 & -e^{i(q_{y}-q_{x})}-e^{-iq_{y}} & -e^{-iq_{x}} \\
 & & 0 & e^{-iq_{y}} & 1 & 0 \\
 &  &  & 0 & 0 & 1 \\
 &  & & & 0 & -e^{-iq_{y}} \\
 -h.c. &  &  & & & 0 \\
\end{array}\right).
\end{equation}
\end{widetext}
We find this $H_{eff}$ does open bulk gap.



\begin{thebibliography}{99}

\bibitem{Anderson73} P. W. Anderson, Mater. Res. Bull. 8, 153 (1973).
\bibitem{Lee08} Patrick. A. Lee, Science, 321, 1306 (2008)
\bibitem{Balents10} L. Balents, Nature (London) 464, 199 (2010).
\bibitem{QSLRMP} Yi Zhou, Kazushi Kanoda and Tai-Kai Ng, Rev. Mod. Phys. 89, 025003 (2017).

\bibitem{Kanoda03} Y. Shimizu, K. Miyagawa, K. Kanoda, M. Maesato and G. Saito, Phys. Rev. Lett. 91, 107001 (2003).
\bibitem{Itou08} T. Itou, A. Oyamada, S. Maegawa, M. Tamura, and R. Kato, Phys. Rev. B 77 104413 (2008).
\bibitem{Shimizu16} Y. Shimizu, T. Hiramatsu, M. Maesato, A. Otsuka, H. Yamochi, A. Ono, M. Itoh, M. Yoshida, M. Takigawa, Y. Yoshida, and G. Saito, Phys. Rev. Lett. 117, 107203 (2016).
\bibitem{Kagome07} J. S. Helton, \emph{et. al.}, Phys. Rev. Lett. 98, 107204 (2007).
\bibitem{Takagi07} Y. Okamoto, M. Nohara, H. Aruga-Katori and H. Takagi, Phys. Rev. Lett. 99, 137207 (2007).

\bibitem{Balents17} L. Savary and L. Balents,  Rep. Prog. Phys. 80, 016502 (2017).




\bibitem{Kitaev} A. Kitaev, Ann. Phys. (Amsterdam) 321, 2 (2006).

\bibitem{TQCRMP} C. Nayak, S. H. Simon, A. Stern, M. Freedman, and S. Das Sarma, Rev. Mod. Phys. 80, 1083 (2008).

\bibitem{Jackeli} G. Jackeli, and G. Khaliullin, Phys. Rev. Lett.  102, 017205 (2009).

\bibitem{yao2007} H. Yao and S. A. Kivelson, Phys. Rev. Lett. 99, 247203 (2007).
\bibitem{yang} S. Yang, D. L. Zhou, and C. P. Sun, Phys. Rev. B 76, 180404 (2007).
\bibitem{baskaran2009} G. Baskaran, G. Santhosh, and R. Shankar, arXiv preprint arXiv:0908.1614 (2009)
\bibitem{tikhonov2010} K. S. Tikhonov and M. V. Feigel’man, Phys. Rev. Lett. 105, 067207 (2010).

\bibitem{si} T. Si and Y. Yu, Nucl. Phys. B 803,428 (2008).
\bibitem{mandal} S. Mandal and N. Surendran, Phys. Rev. B 79, 024426 (2009).
\bibitem{ryu} S. Ryu, Phys. Rev. B 79, 075124 (2009).
\bibitem{Trebst15} M. Hermanns, K. O'Brien, and S. Trebst, Phys. Rev. Lett. 114, 157202 (2015).

\bibitem{wang} F. Wang, Phys. Rev. B 81, 184416 (2010).
\bibitem{yao2011} H. Yao and D.-H. Lee, Phys. Rev. Lett. 107, 087205 (2011).
\bibitem{lai2011} H.-H. Lai and O. I. Motrunich, Phys. Rev. B 83, 155104 (2011).

\bibitem{lee2007} D.-H. Lee, G.-M. Zhang, and T. Xiang, Phys. Rev. Lett. 99, 196805 (2007).
\bibitem{yu2008} Y. Yu and Z. Q. Wang,  Europhysics Letters, 84, 57002 (2008).

\bibitem{yao2009} H. Yao, S.-C. Zhang, and S. A. Kivelson, Phys. Rev. Lett. 102, 217202 (2009).
\bibitem{wu} C. Wu, D. Arovas, and H.-H. Hung, Phys. Rev. B 79, 134427 (2009).
\bibitem{chern} G.-W. Chern, Phys. Rev. B 81, 125134 (2010).
\bibitem{chua} V. Chua, H. Yao, and G. A. Fiete, Phys. Rev. B 83, 180412 (2011).
\bibitem{nakai} R. Nakai, S. Ryu, and A. Furusaki, Phys. Rev. B 85, 155119 (2012).
\bibitem{nussinov2009} Z. Nussinov and G. Ortiz, Phys. Rev. B 79, 214440 (2009).
\bibitem{nussinov2013} Z. Nussinov and Jeroen van den Brink, Rev. Mod. Phys. 87, 1 (2015).

\bibitem{Jordan-Wigner}P. Jordan and E. P. Wigner, Z. Phys. 47, 6319 (1928).

\bibitem{Feng} X. Y. Feng, G. M. Zhang and T. Xiang, Phys. Rev. Lett. 98, 087204 (2007).
\bibitem{Chen} H. D. Chen and J. P. Hu, Phys. Rev. B 76, 193101 (2007).
\bibitem{Chen-Nussinov} H. D. Chen and Z. Nussinov, J. Phys. A 41, 075001 (2008).

\bibitem{edge} Similarly for edge terms \eqref{eq:JWTx} and \eqref{eq:JWTy} become
$\sigma_{L_{x},n_{y},3}^{x}\sigma_{1,n_{y},1}^{z}\sigma_{1,n_{y},2}^{z}\sigma_{1,n_{y},3}^{y}=i\eta_{L_{x},n_{y},3}\eta_{1,n_{y},3}\hat{F}_{n_{y}} $
and $\sigma_{L_{x},n_{y},3}^{x}\sigma_{1,n_{y},1}^{x}\sigma_{L_{x},n_{y}+1,2}^{x}\sigma_{L_{x},n_{y}+1,3}^{y}=-i\hat{D}_{1,n_{y}}\eta_{L_x,ny,3}\eta_{L_x,n_y+1,3}\hat{F}_{n_{y}}$

\bibitem{globalPhi} Indeed, one can define $L_y$ number of global fluxes $\hat{\Phi}_{x}(n_y)=\hat{F}_{n_y}$ with $n_y=1,\cdots,L_y$. But there is only one independent $\hat{\Phi}_{x}$ when local flux configuration $\{\phi_{p}\}$ is fixed.
Similary, there is only one indepedent $\hat{\Phi}_{y}$.

\bibitem{Projection} At first sight, the physical projection should be $\hat{P}^{\prime}=\prod_{n_y}\frac{(1+F_{n_y}\hat{F}_{n_y})}{2}$ to keep all the $\phi_{p}$ on edge plaquettes compatible as well as $\Phi_{x}$ as suggested in Ref.\cite{yao2007}.
But we do not need it in practice, since $\hat{F}_{n_y}\hat{F}_{n_y+1}=\prod^{\prime}_{p}\hat{\phi}_{p}$ and the product of two neighboring edge $\hat{\phi}_p$ commute with the Majorana fermions,
where $\prod^{\prime}$ is over a row of plaquettes between the $n_y$-th and $n_y+1$-th horizontal chain. Indeed, there is only one extra degree of freedom need to project, say, the total fermion number parity $\hat{F}$, as we choose in the main text.

\bibitem{DGS} The preasumed vale $F=1$ in a $\pi$-flux state when both $L_x$ and $L_y$ are even numbers.
On the other hand, for the fermion ground state with $\Phi_x=\Phi_y=1$, the pairing term $\beta_{-\vec{q},\mu} \beta_{\vec{q},\mu}$ vanishes at $\vec{q}=(0,0)$, which gives rise to unpaired fermions at the Fermi level and results in the imcompatible $F=-1$ in filled bands in a finite system.
Thus, this state will be removed by the projection $\hat{P}$. However, $\Phi_x=-1$ or $\Phi_y=-1$ will shift the $\vec{q}=(0,0)$ point by $\pi/L$ on a finite lattice. Such that the pairing term will remain and the other three fermion ground states survive the projection.

\bibitem{CN05} Takahiro Fukui, Yasuhiro Hatsugai, and Hiroshi Suzuki, J. Phys. Soc. Jpn. 74, 1674 (2005).

\bibitem{Miao} J. J. Miao, H. K. Jin, F. C. Zhang and Y. Zhou,  arXiv:1806.10960

\bibitem{MOF} S. R. Batten, S. M. Neville, and D. R. Turner, {\em Coordination Polymers: Design, Analysis and Application}, Royal Society of Chemistry (2009).

\end{thebibliography}
\end{document}